\documentclass[aps,prl,twocolumn,superscriptaddress,longbibliography]{revtex4-2}
\usepackage{amsmath,amssymb}
\usepackage[pdftex]{hyperref,graphicx}
\hypersetup{colorlinks = true, urlcolor = blue, linkcolor = blue, citecolor = blue}
\usepackage{physics}
\usepackage{xcolor}
\usepackage{bm}
\usepackage{pdfpages}

\newcommand{\comment}[1]{{}}

\makeatletter
\AtBeginDocument{\let\LS@rot\@undefined}
\makeatother

\begin{document}
\title{Temperature dependent resistivity in the doped two dimensional metallic phase of mTMD bilayers}
\author{Seongjin Ahn}
\affiliation{Condensed Matter Theory Center and Joint Quantum Institute, Department of Physics, University of Maryland, College Park, Maryland 20742, USA}
\author{Sankar Das Sarma}
\affiliation{Condensed Matter Theory Center and Joint Quantum Institute, Department of Physics, University of Maryland, College Park, Maryland 20742, USA}

\begin{abstract}
Two recent experiments from Cornell and Columbia have reported insulator-to-metal transitions in two-dimensional (2D) moiré transition metal dichalcogenides (mTMD) induced by doping around half-filling, where the system is a Mott insulator.  In the current work, we consider the temperature dependent resistivity of this metallic phase in the doped situation away from half-filling, arguing that it arises from the strongly temperature dependent 2D Friedel oscillations (i.e. finite momentum screening) associated with random quenched charged impurities, leading to the observed strongly increasing linear-in-$T$ resistivity in the metallic phase.  Our theory appears to account for the temperature-dependent metallic resistivity for doping around half-filling of the effective moiré TMD band, showing that temperature-dependent screened Coulomb disorder is an essential ingredient of doped 2D mTMD physics.
\end{abstract}

\maketitle
{\em Introduction.}---  
The Mott insulator (MI) is a strongly correlated electronic phase, where a half-filled (generally, narrow) band, instead of being a metal as simple electron counting implies, is rendered an insulator because of strong interactions suppressing the itinerant electronic hopping so as to minimize interaction  effects.  This happens when the dimensionless parameter $U/t \gg 1$, where $U(t)$ is the effective interaction (hopping) strength, and narrow band systems with small $t$ are the prime candidates for the MI phase.  Understanding all aspects of MI physics is one of the major thrusts in condensed matter physics, and has been that way for more than 50 years, but the problem is a difficult non-perturbative strong-coupling many-electron problem not amenable to easy theoretical (or even numerical) solutions. Recent experiments in moiré homo- and hetero-bilayer TMD (mTMD) structures have reported the observation of strongly correlated insulating phases at various rational fillings including not only half-filling (i.e., 1 spinful electron per lattice site), but also $1/4$, $2/3$, etc. rational fillings \cite{liContinuousMottTransition2021b, ghiottoQuantumCriticalityTwisted2021a, xuCorrelatedInsulatingStates2020a, huangCorrelatedInsulatingStates2021, wangCorrelatedElectronicPhases2020a}. 
These observations have been theoretically interpreted as the manifestation of MI-like correlated insulator phases which also naturally manifest charge density and spin density waves \cite{wuHubbardModelPhysics2018, wuTopologicalInsulatorsTwisted2019, panQuantumPhaseDiagram2020, panBandTopologyHubbard2020, panInteractionDrivenFillingInducedMetalInsulator2021a, panInteractionRangeTemperature2022}. 
The current theoretical work is focused not on the half-filled MI phase, but on the situation where the mTMD system is doped slightly away from half-filling where it should be a metal \cite{panInteractionDrivenFillingInducedMetalInsulator2021a}.  In general, a doped MI should always be a metal, superficially similar to a doped semiconductor although the origin of the undoped MI phase is qualitatively different from the single-particle band picture applicable to undoped semiconductors.


Recent experiments from Cornell \cite{liContinuousMottTransition2021b} 
and Columbia \cite{ghiottoQuantumCriticalityTwisted2021a} 
report the observations of 2D metal-insulator transitions (MIT), where doping the half-filled correlated MI state in both homobilayer \cite{ghiottoQuantumCriticalityTwisted2021a} and heterobilayer \cite{liContinuousMottTransition2021b} mTMD systems eventually at some finite critical doping, but not for infinitesimal doping, converts the system to a metal with a finite $T=0$ conductivity (of course, the MI by definition has zero $T=0$ conductivity). This doping-induced MIT around half-filling is the main subject of the current theoretical work. In particular, the experiments report strong linear-in-$T$ resistivity (with resistivity increasing linearly with increasing $T$) in the metallic phase above the critical doping in contrast to their observed $T^2$ resistivity \cite{liContinuousMottTransition2021b, ghiottoQuantumCriticalityTwisted2021a} in the metallic phase at half-filling (obtained by tuning an applied external electric field presumably tuning the effective $U/t$ so as to induce the Mott transition, we do not discuss the strong correlation driven MIT at half-filling  since it is outside the scope of our theory). Presumably the $T^2$ metallic temperature dependence at half-filling beyond the MIT arises from the usual Fermi liquid behavior of umklapp electron-electron scattering, but the linear-in-$T$ temperature dependence of the doping-induced metallic state above the critical doping is anomalous since the expectation is that this $T$-dependence should also be the usual Fermi liquid $T^2$ behavior as the finite-doping metal slightly away from half-filling should be a normal metal. Based on this observed linear-in-$T$ resistivity behavior, the experimental finite doping metallic phase around half-filling has been claimed to be a putative non-Fermi liquid arising from some type of a continuous quantum critical MIT, which generally should not happen in 2D in the presence of any disorder by virtue of the scaling theory of localization. We mention that there is no known (or proposed) quantum critical point in the doped system around half-filling leading to a linear-in-T resistivity. In the current work, we provide a physically-motivated explanation for the linearly temperature-dependent resistivity in the doped metallic case, arising from screened Coulomb disorder due to the inevitable presence of unintentional quenched random charged impurities in the TMD environment. This same Coulomb disorder is also responsible for producing the apparent 2D MIT crossover behavior itself (i.e. producing the localized insulating phase around half-filling) in the doped samples as we have argued recently \cite{ahnDisorderinducedTwodimensionalMetalinsulator2022}.

{\em Theory.}---  
The resistivity can be calculated using the usual Drude formula
\begin{equation}
    \rho(T) = \frac{m}{n e^2 \tau (T)}.
    \label{eq:finite_temperature_resistivity}
\end{equation}
where $m=0.45$ (in units of free-electron mass) is the appropriate effective mass \cite{Mak} and $n$ is the mobile carrier density. Since the mTMD systems are slightly doped away from half-filling, only the fraction of electrons occupying the upper Hubbard band can participate in the metallic transport, and thus throughout the paper we set $n=n_\mathrm{T}-n_\mathrm{M}$ as the mobile carrier density in the metallic phase, where $n_\mathrm{T}$ is the total carrier density, and  $n_M=5\times10^{12}\mathrm{cm}^{-2}$ is the moiré density corresponding to half filling.
$\tau (T)$ is the temperature-dependent scattering time within the Boltzmann transport theory given by
\begin{equation}
    \tau(T)=\frac{  \int d\varepsilon_\mathrm{\bm k}  \tau(\mathrm{\varepsilon_\mathrm{\bm k}}) \varepsilon_\mathrm{\bm k}\left(-\frac{ \partial f(\varepsilon_\mathrm{\bm k})}{\partial\varepsilon_\mathrm{\bm k}}\right)  }
    {\int d \varepsilon_\mathrm{\bm k} \varepsilon_\mathrm{\bm k} \left(-\frac{ \partial f(\varepsilon_\mathrm{\bm k})}{\partial\varepsilon_\mathrm{\bm k}}\right) },
    \label{eq:tau_finite_T}
\end{equation} 
where $\tau(\mathrm{\varepsilon_\mathrm{\bm k}})$ denotes the zero-temperature transport time limited by screened charged disorder, which we calculate using the leading-order Boltzmann transport theory:
\begin{equation} 
    \frac{1}{\tau({\varepsilon_\mathrm{\bm k}})}=\frac{2\pi n_i}{\hbar} 
    \sum_\mathrm{\bm k'}
    \left|u_\mathrm{i}(\bm k - \bm k')\right|^2
    (1-\cos{ \theta})\delta(\epsilon_\mathrm{\bm k}-\epsilon_\mathrm{\bm k'}),
    \label{eq:tau_zero_T}
\end{equation} 
where $n_i$ is the background unintentional random charged impurity density, $u_\mathrm{i}(\bm q)=2\pi e^2/ \kappa q \epsilon(q,T)$ is the screened Coulomb scattering potential between a charged impurity and an electron with $\epsilon(q,T)=1+(2\pi e^2/ \kappa q)\Pi(q,T)$ denoting the RPA finite-temperature and finite-momentum dielectric function and $\kappa=5$ the background lattice dielectric constant for mTMD. The finite temperature polarizability $\Pi(q,T)$ can be expressed as an integral of the zero-temperature polarizability:
\begin{equation}
    \Pi(q,T)=\int_\mathrm{0}^{\infty} d\epsilon \frac{\Pi(q,T=0)|_{\epsilon_\mathrm{F}=\epsilon}}{4k_\mathrm{B}T\cosh^2{\frac{\epsilon-\mu(T)}{2k_\mathrm{B}T}}}.
    \label{eq:pol_finite_T}
\end{equation}
where
 \begin{equation}
    \Pi(q,T=0)=
		\frac{m}{\pi\hbar^2}\left[1 - \Theta(q-2k_\mathrm{F})\frac{\sqrt{q^2- 4k^2_\mathrm{F} }}{q}\right],
    \label{eq:pol_zero_T}
\end{equation}
is the zero-temperature 2D polarizability \cite{sternPolarizabilityTwoDimensionalElectron1967}. Here $\mu(T)$ and $k_\mathrm{F}$ are the chemical potential and the Fermi wavevector, respectively. With Eqs.~(\ref{eq:finite_temperature_resistivity})-(\ref{eq:pol_zero_T}), the low- and high-temperature resistivity can be asymptotically expanded as \cite{dassarmaScreeningTransport2D2015}
\begin{equation}
    \begin{aligned}
        \rho(T \ll T_\mathrm{F}) &\approx \rho_0\left[\frac{2x}{1+x}\left(\frac{T}{T_\mathrm{F}}\right) +
                        \frac{2.646x}{(1+x)^2} \left(\frac{T}{T_\mathrm{F}}\right)^{3/2} \right] \\
        \rho(T \gg T_\mathrm{F}) &\approx  \rho_1 \left(\frac{T_\mathrm{F}}{T}\right) \left[1-\frac{3\sqrt x}{4}\left(\frac{T_\mathrm{F}}{T}\right)^{3/2} \right] 
    \end{aligned}
    \label{eq:resistivity_asymptotic_low_and_large_T}
\end{equation}
where $q_\mathrm{TF}= 2 m e^2/\kappa\hbar^2$ is the Thosmas-Fermi wavevector, $\rho_0=\rho(T=0)$ is the zero-temperature residual resistivity induced by impurity scattering, $\rho_1=(h/e^2) (n_i/n) (\pi x^2/2)$ and $x=q_\mathrm{TF}/2k_\mathrm{F}$.
We note and emphasize that the temperature dependent resistivity explicitly analytically follows the low-$T$ behavior $\rho (T)/\rho_0 = 1 + A(T/T_\mathrm{F})$, where $A$ is a doping dependent constant, precisely as  experiments observe \cite{liContinuousMottTransition2021b, ghiottoQuantumCriticalityTwisted2021a}. 
This linearity, which violates the well-known Sommerfeld expansion, arises from the $2k_\mathrm{F}$-anomaly of the 2D screening function, where the temperature effect in suppressing $2k_\mathrm{F}$-screening is strong going as $\mathcal{O}(T^{1/2})$ rather than the exponentially weak $T$-dependence manifesting at long wavelength screening \cite{dassarmaScreeningTransport2D2015}. 
Thus, the linear-in-$T$ behavior ultimately arises from the anomalous $T$-dependence of 2D Friedel oscillations, leading to the nonanalytic $T$-dependence of screening at $2k_\mathrm{F}$. This physics could qualitatively be construed as the high-temperature analog of the 2D Altshuler-Aronov effect.

We emphasize that this linearly T dependent metallic resistivity in our theory (and presumably as observed experimentally in \cite{liContinuousMottTransition2021b, ghiottoQuantumCriticalityTwisted2021a}) does not arise from any $T=0$ quantum criticality and does not imply any non-Fermi liquid behavior. It is simply a nonanalytic anomalous finite-T property of the 2D Fermi surface, where the temperature dependence deviates qualitatively from that given by Sommerfeld expansion arising from the 2$k_\mathrm{F}$-anomaly \cite{sternCalculatedTemperatureDependence1980, dassarmaTheoryFinitetemperatureScreening1986, zalaInteractionCorrectionsIntermediate2001, buterakosPresenceAbsenceTwodimensional2021}.
We also note that any phonon scattering induced $T$-dependence is irrelevant in the temperature range of interest, because the electron-phonon coupling is very weak in TMD materials \cite{Mak} and also because the phonon-induced temperature dependence is negligible, going as $T^4$ in this low temperature range well below the Bloch-Gruneisen temperature.

\begin{figure}[!htb]
  \centering
  \includegraphics[width=\linewidth]{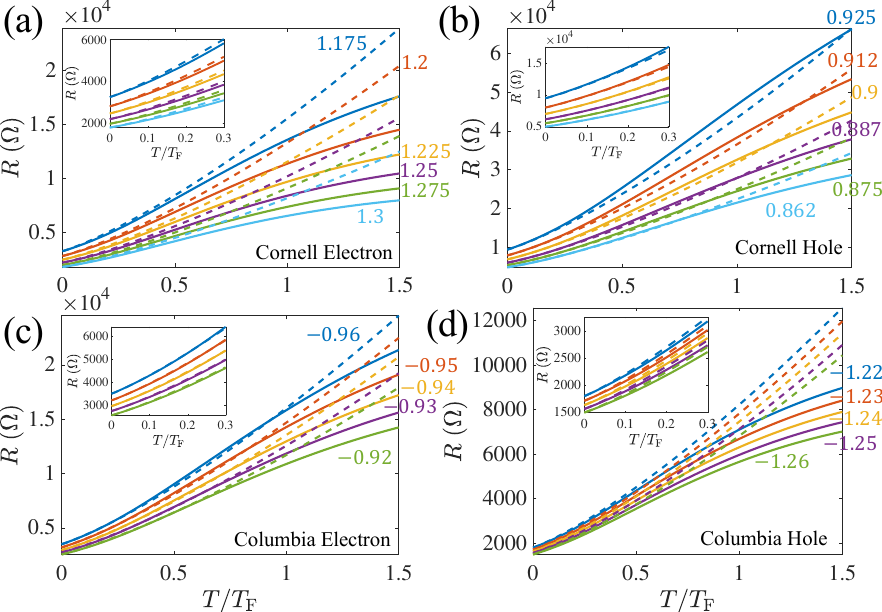}
  \caption{Numerically calculated resistivity (solid lines) as a function of $T$ along with the analytical asymptotic results (dashed lines) for doped metallic mTMD samples from the (a), (b) Cornell and (c), (d) Columbia groups. Each resistivity curve in a different color corresponds to a different filling factor, which is indicated by a colored number in the figure. The insets present a zoom-in of the low-temperature regime, where the full numerical and analytical results are in good agreement exhibiting a linear $T$ dependence at $T\ll T_\mathrm{F}$. For calculations, we set the impurity densities $n_\mathrm{i}=$ (a) $8.71$, (b) $10.1$, (c) $6.2$ and (d) $7.8\times 10^{10}\mathrm{cm}^{-2}$, which are estimated by fitting the lowest-temperature measured resistivity \cite{liContinuousMottTransition2021b, ghiottoQuantumCriticalityTwisted2021a} to the leading-order Boltzmann theory as a function of carrier density \cite{ahnDisorderinducedTwodimensionalMetalinsulator2022}.  The estimated impurity density is reasonable for TMD materials \cite{Mak}. $T_\mathrm{F}$ in each figure is given by $T_\mathrm{F}= \left|f-1 \right| 308.7$K for (a) and (b), and $T_\mathrm{F}= \left|f+1 \right| 308.7$K for (c) and (d), where $f$ is the filling factor.   }  
  \label{fig:1}
\end{figure}

We present in the next section our numerical results for the full transport theory using mTMD parameters from Refs.~\cite{liContinuousMottTransition2021b, ghiottoQuantumCriticalityTwisted2021a}. 
We note that, while $\rho_0$ depends on the impurity parameters, the temperature dependence itself is independent of disorder details in the metallic phase.

{\em Results.}---  
In Fig.~\ref{fig:1} we show the calculated $T$-dependence of the doped metallic mTMD resistivity based on our screened disorder scattering as described above.  
The results shown in Fig.~\ref{fig:1} are in good qualitative and semi-quantitative agreement with the experimental results of \cite{ghiottoQuantumCriticalityTwisted2021a, liContinuousMottTransition2021b, Mak2}. 
In particular, the three most significant salient features of the experimental results are reproduced well in the theory: (1) A linear $T$-dependence in the resistivity in the leading order at the lowest temperatures; (2) the $T$-dependence becoming stronger with decreasing doping density approaching the MIT critical density from the metallic side; (3) the $T$-dependence could be by as much as a factor of 2-6 for a modest temperature increase of $\sim20$K as observed experimentally. We emphasize that our Boltzmann theory based transport theory becomes progressively worse quantitatively as the MIT is approached, remaining qualitatively (but not quantitatively valid) near the effective MIT itself.  We do not provide a direct comparison with the actual experimental data because it is not meaningful to do so because of (1) the approximate zeroth order nature of our theory, and (2) the experimental results show significant sample-to-sample variations, implying that only semi-quantitative and qualitative comparison between theory and experiment is sensible at this point. 
The most significant success of the theory is in providing a natural explanation for the linear-in-T behavior and its doping dependence and the fact that the required impurity density, the only free parameter in the theory providing the correct quantitative resistivity scale agreeing with the experiment, is reasonable.

We must emphasize that the linear-in-$T$ resistivity extending to low temperatures in our theory is not connected with any non-Fermi liquid behavior whatsoever, the metallic system remains a Fermi liquid here everywhere since the linear-in-$T$  transport scattering rate is simply a momentum relaxation rate relevant for current relaxation, and has nothing to do with any dynamical imaginary self-energy varying as linear-in-$T$. Any non-Fermi-liquid behavior requires the inelastic scattering rate to go as (at least) linear-in-$T$ and the $T$-dependence of the momentum relaxation by itself is irrelevant.

\begin{figure}[!htb]
  \centering
  \includegraphics[width=\linewidth]{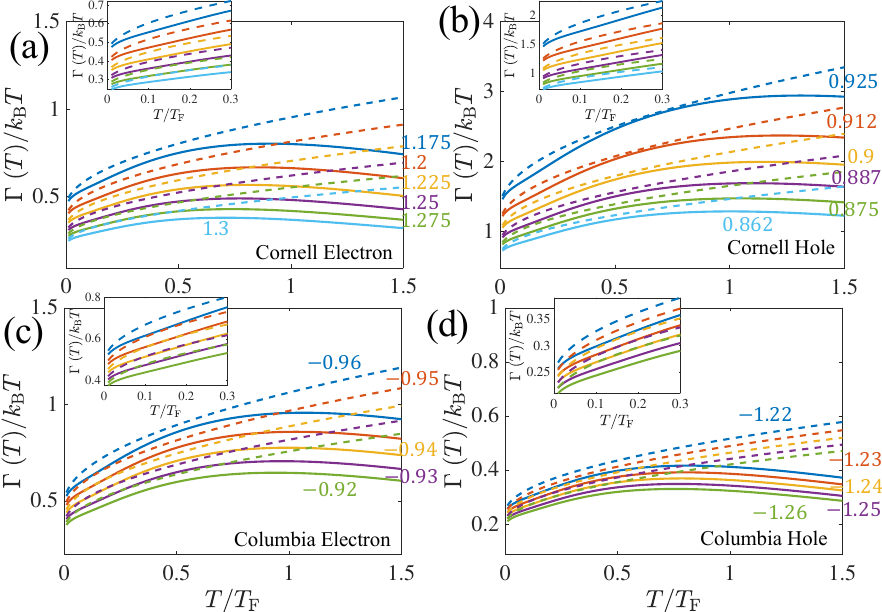}
  \caption{The calculated scattering rate $\Gamma=\hbar/\tau$ scaled by temperature as a function of $T/T_\mathrm{F}$ for the results of Fig.~\ref{fig:1}, indicating the super-Planckian ($\Gamma/k_B T>1$) in Fig.~\ref{fig:2}(b) arising from temperature dependent screened disorder scattering. Note that this figure shows only the $T$-dependent scattering subtracting out the $T=0$ disorder scattering rate.}  
  \label{fig:2}
\end{figure}

In Fig.~\ref{fig:2} we plot the calculated finite-$T$ scattering rate, defined by $\hbar/\tau(T)$ where $1/\tau (T) = m[ \rho(T) - \rho (T=0)] /(ne^2)$, as a function of $T$ for different carrier densities $n$, in order to emphasize the `Planckian' nature \cite{hartnollPlanckianDissipationMetals2021, ahnPlanckianProperties2D2022} of transport in doped mTMD layers although the Planckian behavior here is not associated with any non-Fermi liquid property or hidden quantum criticality, arising specifically from the 2D Fermi liquid $2k_\mathrm{F}$-anomaly.  The fact that $\hbar/\tau>T$ in Fig.~\ref{fig:2}(b) indicates the strange  metallic super-Planckian behavior for the Cornell doped hole data, whereas the other three sets of data are sub-Planckian. 


\begin{figure}[!htb]
  \centering
  \includegraphics[width=\linewidth]{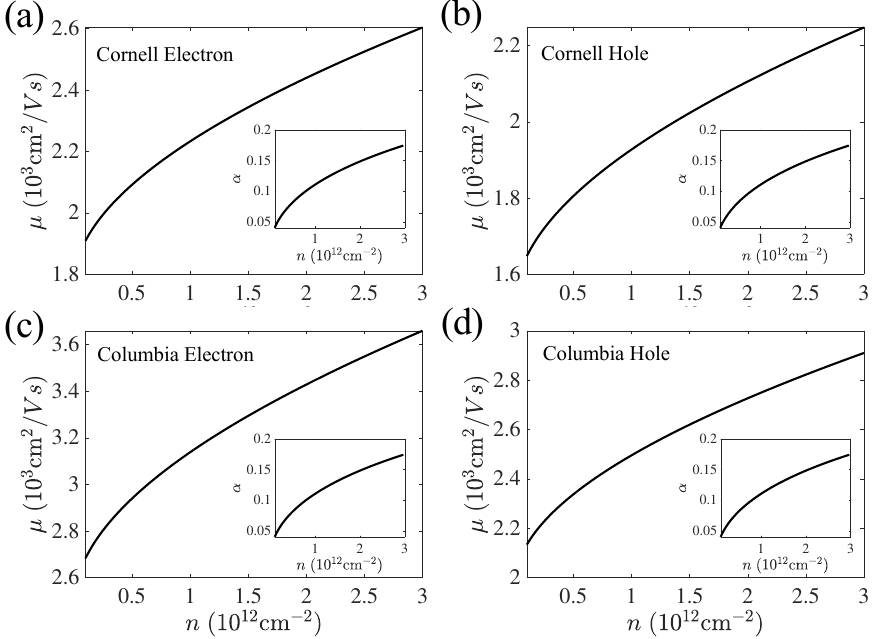}
  \caption{Calculated mobility for each mTMD sample obtained by fitting the experimental resistivity measured at the lowest temperature to the leading-order Boltzmann theory \cite{ahnDisorderinducedTwodimensionalMetalinsulator2022}. The insets show the density-dependent power-law exponent as a function of $n$ calculated numerically using $\alpha=\partial\ln{\mu}/\partial \ln{n}$. We note that the mobilities ($\sim 10^3 \mathrm{cm}^2/Vs$) are rather low indicating the dominant role of disorder, and the exponent $\alpha$ being small ($\ll 1$) indicates the strong role of screening.}  
  \label{fig:3}
\end{figure}

For the sake of completeness, we have also calculated the carrier density dependent mobility, $\mu$, indicating the sample quality for both Cornell and Columbia doped samples using the one-parameter impurity density fit, which is fixed by comparing the measured resistivity with the theory just at one doping density at the lowest temperature. The mobility $\mu$ is defined by
\begin{equation}
    \mu= \frac{1}{n e \rho},    
\end{equation}
where $\rho$ is the resistivity at the lowest temperature (i.e., $\rho_0$ in the theory). These calculated results as well as the power law fits to $\mu \sim n^{\alpha}$, where $\alpha$ is the density exponent of $\mu(n)$, are shown in Fig.~\ref{fig:3} for Cornell and Columbia experiments, respectively, both on the electron and hole doping sides. For these calculations, we set the impurity densities $n_\mathrm{i}$= (a) $8.71$, (b) $10.1$, (c) $6.2$ and (d) $7.8\times 10^{10}\mathrm{cm}^{-2}$, which are estimated by fitting the density-dependent experimental resistivity to the leading-order Boltzmann theory \cite{ahnDisorderinducedTwodimensionalMetalinsulator2022}.
We note that the best-fit impurity density in all cases is $n_\mathrm{i} \sim 5\times10^{10} \mathrm{cm}^{-2}$, which is consistent with the known quality of these TMD materials \cite{Mak}. There is a small difference between the impurity content on the electron/hole doping sides consistent with the experimentally observed asymmetry in the experiments on the two doping sides. 
Our calculated mobility and its density dependence both agree well with the experiments.
The fact that our theory accounts for the measured sample mobility and its doping dependence naturally with very reasonable choices for the sample disorder provide strong support for our disorder model of the mTMD transport physics in doped samples.

We emphasize that the increasing mobility with increasing doping is a direct manifestation of the effect of screening of the Coulomb disorder by the doped carriers, and cannot be explained by the Mott-Hubbard type strong correlation model at all.  In particular, the limiting theoretical values of $\alpha$ can be deduced from our theory analytically with $\alpha(x\ll1) = 1.5$ as appropriate for unscreened disorder and $\alpha (x \gg 1) = 0$ as appropriate for totally screened short-range disorder (with $x=q_\mathrm{TF}/2k_\mathrm{F}$) \cite{dassarmaUniversalDensityScaling2013}. The fact that $\alpha<1$ throughout in Fig.~\ref{fig:3} indicates that mTMD systems are strong screening systems (by virtue of its large effective mass), and the low mobility value ($\sim10^3\mathrm{cm}^2/Vs$) is simply a manifestation of the highly disordered nature of the current mTMD samples. Improvement in sample quality should lead to much wider metallic regimes on both electron/hole sides of future mTMD samples \cite{comm}. 

{\em Conclusion.}---  
We provide a plausible theory for the strong temperature dependence of the reported 2D metallic resistivity of mTMD bilayers doped away from half-filling.  Our theory explains the observed strong temperature dependence of the doped metal as arising from the screened Coulomb disorder, and not from any unknown quantum criticality or mysterious non-Fermi liquid properties.  Our microscopic disorder-based transport theory is in agreement with the experimental finding of a linear-in-$T$ resistivity down to the lowest experimental temperatures as well as the strong $T$-dependence, along with a Planckian behavior with the scattering rate exceeding temperature itself.  One obvious implication of our disorder-based theory is that cleaner (dirtier) systems with less (more) disorder should manifest a larger (smaller) metallic regime with a smaller (larger) critical doping density for the observed 2D MIT. Our work suggests that doping-induced effective MIT in mTMD bilayers may be arising from Coulomb disorder and not from strong correlation.

\section{Acknowledgement} \label{sec:acknowledgement}
We thank K. Mak, C. Dean, and A. Pasupathy for helpful discussions on the experimental data and for providing us unpublished results \cite{Mak2}. 
This work is supported by the Laboratory for Physical Sciences.


\begin{thebibliography}{24}%
\makeatletter
\providecommand \@ifxundefined [1]{%
 \@ifx{#1\undefined}
}%
\providecommand \@ifnum [1]{%
 \ifnum #1\expandafter \@firstoftwo
 \else \expandafter \@secondoftwo
 \fi
}%
\providecommand \@ifx [1]{%
 \ifx #1\expandafter \@firstoftwo
 \else \expandafter \@secondoftwo
 \fi
}%
\providecommand \natexlab [1]{#1}%
\providecommand \enquote  [1]{``#1''}%
\providecommand \bibnamefont  [1]{#1}%
\providecommand \bibfnamefont [1]{#1}%
\providecommand \citenamefont [1]{#1}%
\providecommand \href@noop [0]{\@secondoftwo}%
\providecommand \href [0]{\begingroup \@sanitize@url \@href}%
\providecommand \@href[1]{\@@startlink{#1}\@@href}%
\providecommand \@@href[1]{\endgroup#1\@@endlink}%
\providecommand \@sanitize@url [0]{\catcode `\\12\catcode `\$12\catcode
  `\&12\catcode `\#12\catcode `\^12\catcode `\_12\catcode `\%12\relax}%
\providecommand \@@startlink[1]{}%
\providecommand \@@endlink[0]{}%
\providecommand \url  [0]{\begingroup\@sanitize@url \@url }%
\providecommand \@url [1]{\endgroup\@href {#1}{\urlprefix }}%
\providecommand \urlprefix  [0]{URL }%
\providecommand \Eprint [0]{\href }%
\providecommand \doibase [0]{https://doi.org/}%
\providecommand \selectlanguage [0]{\@gobble}%
\providecommand \bibinfo  [0]{\@secondoftwo}%
\providecommand \bibfield  [0]{\@secondoftwo}%
\providecommand \translation [1]{[#1]}%
\providecommand \BibitemOpen [0]{}%
\providecommand \bibitemStop [0]{}%
\providecommand \bibitemNoStop [0]{.\EOS\space}%
\providecommand \EOS [0]{\spacefactor3000\relax}%
\providecommand \BibitemShut  [1]{\csname bibitem#1\endcsname}%
\let\auto@bib@innerbib\@empty
\bibitem [{\citenamefont {Li}\ \emph {et~al.}(2021)\citenamefont {Li},
  \citenamefont {Jiang}, \citenamefont {Li}, \citenamefont {Zhang},
  \citenamefont {Kang}, \citenamefont {Zhu}, \citenamefont {Watanabe},
  \citenamefont {Taniguchi}, \citenamefont {Chowdhury}, \citenamefont {Fu},
  \citenamefont {Shan},\ and\ \citenamefont
  {Mak}}]{liContinuousMottTransition2021b}%
  \BibitemOpen
  \bibfield  {author} {\bibinfo {author} {\bibfnamefont {T.}~\bibnamefont
  {Li}}, \bibinfo {author} {\bibfnamefont {S.}~\bibnamefont {Jiang}}, \bibinfo
  {author} {\bibfnamefont {L.}~\bibnamefont {Li}}, \bibinfo {author}
  {\bibfnamefont {Y.}~\bibnamefont {Zhang}}, \bibinfo {author} {\bibfnamefont
  {K.}~\bibnamefont {Kang}}, \bibinfo {author} {\bibfnamefont {J.}~\bibnamefont
  {Zhu}}, \bibinfo {author} {\bibfnamefont {K.}~\bibnamefont {Watanabe}},
  \bibinfo {author} {\bibfnamefont {T.}~\bibnamefont {Taniguchi}}, \bibinfo
  {author} {\bibfnamefont {D.}~\bibnamefont {Chowdhury}}, \bibinfo {author}
  {\bibfnamefont {L.}~\bibnamefont {Fu}}, \bibinfo {author} {\bibfnamefont
  {J.}~\bibnamefont {Shan}},\ and\ \bibinfo {author} {\bibfnamefont {K.~F.}\
  \bibnamefont {Mak}},\ }\bibfield  {title} {\bibinfo {title} {Continuous
  {{Mott}} transition in semiconductor moir\'e superlattices},\ }\href
  {https://doi.org/10.1038/s41586-021-03853-0} {\bibfield  {journal} {\bibinfo
  {journal} {Nature}\ }\textbf {\bibinfo {volume} {597}},\ \bibinfo {pages}
  {350} (\bibinfo {year} {2021})}\BibitemShut {NoStop}%
\bibitem [{\citenamefont {Ghiotto}\ \emph {et~al.}(2021)\citenamefont
  {Ghiotto}, \citenamefont {Shih}, \citenamefont {Pereira}, \citenamefont
  {Rhodes}, \citenamefont {Kim}, \citenamefont {Zang}, \citenamefont {Millis},
  \citenamefont {Watanabe}, \citenamefont {Taniguchi}, \citenamefont {Hone},
  \citenamefont {Wang}, \citenamefont {Dean},\ and\ \citenamefont
  {Pasupathy}}]{ghiottoQuantumCriticalityTwisted2021a}%
  \BibitemOpen
  \bibfield  {author} {\bibinfo {author} {\bibfnamefont {A.}~\bibnamefont
  {Ghiotto}}, \bibinfo {author} {\bibfnamefont {E.-M.}\ \bibnamefont {Shih}},
  \bibinfo {author} {\bibfnamefont {G.~S. S.~G.}\ \bibnamefont {Pereira}},
  \bibinfo {author} {\bibfnamefont {D.~A.}\ \bibnamefont {Rhodes}}, \bibinfo
  {author} {\bibfnamefont {B.}~\bibnamefont {Kim}}, \bibinfo {author}
  {\bibfnamefont {J.}~\bibnamefont {Zang}}, \bibinfo {author} {\bibfnamefont
  {A.~J.}\ \bibnamefont {Millis}}, \bibinfo {author} {\bibfnamefont
  {K.}~\bibnamefont {Watanabe}}, \bibinfo {author} {\bibfnamefont
  {T.}~\bibnamefont {Taniguchi}}, \bibinfo {author} {\bibfnamefont {J.~C.}\
  \bibnamefont {Hone}}, \bibinfo {author} {\bibfnamefont {L.}~\bibnamefont
  {Wang}}, \bibinfo {author} {\bibfnamefont {C.~R.}\ \bibnamefont {Dean}},\
  and\ \bibinfo {author} {\bibfnamefont {A.~N.}\ \bibnamefont {Pasupathy}},\
  }\bibfield  {title} {\bibinfo {title} {Quantum criticality in twisted
  transition metal dichalcogenides},\ }\href
  {https://doi.org/10.1038/s41586-021-03815-6} {\bibfield  {journal} {\bibinfo
  {journal} {Nature}\ }\textbf {\bibinfo {volume} {597}},\ \bibinfo {pages}
  {345} (\bibinfo {year} {2021})}\BibitemShut {NoStop}%
\bibitem [{\citenamefont {Xu}\ \emph {et~al.}(2020)\citenamefont {Xu},
  \citenamefont {Liu}, \citenamefont {Rhodes}, \citenamefont {Watanabe},
  \citenamefont {Taniguchi}, \citenamefont {Hone}, \citenamefont {Elser},
  \citenamefont {Mak},\ and\ \citenamefont
  {Shan}}]{xuCorrelatedInsulatingStates2020a}%
  \BibitemOpen
  \bibfield  {author} {\bibinfo {author} {\bibfnamefont {Y.}~\bibnamefont
  {Xu}}, \bibinfo {author} {\bibfnamefont {S.}~\bibnamefont {Liu}}, \bibinfo
  {author} {\bibfnamefont {D.~A.}\ \bibnamefont {Rhodes}}, \bibinfo {author}
  {\bibfnamefont {K.}~\bibnamefont {Watanabe}}, \bibinfo {author}
  {\bibfnamefont {T.}~\bibnamefont {Taniguchi}}, \bibinfo {author}
  {\bibfnamefont {J.}~\bibnamefont {Hone}}, \bibinfo {author} {\bibfnamefont
  {V.}~\bibnamefont {Elser}}, \bibinfo {author} {\bibfnamefont {K.~F.}\
  \bibnamefont {Mak}},\ and\ \bibinfo {author} {\bibfnamefont {J.}~\bibnamefont
  {Shan}},\ }\bibfield  {title} {\bibinfo {title} {Correlated insulating states
  at fractional fillings of moir\'e superlattices},\ }\href
  {https://doi.org/10.1038/s41586-020-2868-6} {\bibfield  {journal} {\bibinfo
  {journal} {Nature}\ }\textbf {\bibinfo {volume} {587}},\ \bibinfo {pages}
  {214} (\bibinfo {year} {2020})}\BibitemShut {NoStop}%
\bibitem [{\citenamefont {Huang}\ \emph {et~al.}(2021)\citenamefont {Huang},
  \citenamefont {Wang}, \citenamefont {Miao}, \citenamefont {Wang},
  \citenamefont {Li}, \citenamefont {Lian}, \citenamefont {Taniguchi},
  \citenamefont {Watanabe}, \citenamefont {Okamoto}, \citenamefont {Xiao},
  \citenamefont {Shi},\ and\ \citenamefont
  {Cui}}]{huangCorrelatedInsulatingStates2021}%
  \BibitemOpen
  \bibfield  {author} {\bibinfo {author} {\bibfnamefont {X.}~\bibnamefont
  {Huang}}, \bibinfo {author} {\bibfnamefont {T.}~\bibnamefont {Wang}},
  \bibinfo {author} {\bibfnamefont {S.}~\bibnamefont {Miao}}, \bibinfo {author}
  {\bibfnamefont {C.}~\bibnamefont {Wang}}, \bibinfo {author} {\bibfnamefont
  {Z.}~\bibnamefont {Li}}, \bibinfo {author} {\bibfnamefont {Z.}~\bibnamefont
  {Lian}}, \bibinfo {author} {\bibfnamefont {T.}~\bibnamefont {Taniguchi}},
  \bibinfo {author} {\bibfnamefont {K.}~\bibnamefont {Watanabe}}, \bibinfo
  {author} {\bibfnamefont {S.}~\bibnamefont {Okamoto}}, \bibinfo {author}
  {\bibfnamefont {D.}~\bibnamefont {Xiao}}, \bibinfo {author} {\bibfnamefont
  {S.-F.}\ \bibnamefont {Shi}},\ and\ \bibinfo {author} {\bibfnamefont {Y.-T.}\
  \bibnamefont {Cui}},\ }\bibfield  {title} {\bibinfo {title} {Correlated
  insulating states at fractional fillings of the {{WS2}}/{{WSe2}} moir\'e
  lattice},\ }\href {https://doi.org/10.1038/s41567-021-01171-w} {\bibfield
  {journal} {\bibinfo  {journal} {Nat. Phys.}\ }\textbf {\bibinfo {volume}
  {17}},\ \bibinfo {pages} {715} (\bibinfo {year} {2021})}\BibitemShut
  {NoStop}%
\bibitem [{\citenamefont {Wang}\ \emph {et~al.}(2020)\citenamefont {Wang},
  \citenamefont {Shih}, \citenamefont {Ghiotto}, \citenamefont {Xian},
  \citenamefont {Rhodes}, \citenamefont {Tan}, \citenamefont {Claassen},
  \citenamefont {Kennes}, \citenamefont {Bai}, \citenamefont {Kim},
  \citenamefont {Watanabe}, \citenamefont {Taniguchi}, \citenamefont {Zhu},
  \citenamefont {Hone}, \citenamefont {Rubio}, \citenamefont {Pasupathy},\ and\
  \citenamefont {Dean}}]{wangCorrelatedElectronicPhases2020a}%
  \BibitemOpen
  \bibfield  {author} {\bibinfo {author} {\bibfnamefont {L.}~\bibnamefont
  {Wang}}, \bibinfo {author} {\bibfnamefont {E.-M.}\ \bibnamefont {Shih}},
  \bibinfo {author} {\bibfnamefont {A.}~\bibnamefont {Ghiotto}}, \bibinfo
  {author} {\bibfnamefont {L.}~\bibnamefont {Xian}}, \bibinfo {author}
  {\bibfnamefont {D.~A.}\ \bibnamefont {Rhodes}}, \bibinfo {author}
  {\bibfnamefont {C.}~\bibnamefont {Tan}}, \bibinfo {author} {\bibfnamefont
  {M.}~\bibnamefont {Claassen}}, \bibinfo {author} {\bibfnamefont {D.~M.}\
  \bibnamefont {Kennes}}, \bibinfo {author} {\bibfnamefont {Y.}~\bibnamefont
  {Bai}}, \bibinfo {author} {\bibfnamefont {B.}~\bibnamefont {Kim}}, \bibinfo
  {author} {\bibfnamefont {K.}~\bibnamefont {Watanabe}}, \bibinfo {author}
  {\bibfnamefont {T.}~\bibnamefont {Taniguchi}}, \bibinfo {author}
  {\bibfnamefont {X.}~\bibnamefont {Zhu}}, \bibinfo {author} {\bibfnamefont
  {J.}~\bibnamefont {Hone}}, \bibinfo {author} {\bibfnamefont {A.}~\bibnamefont
  {Rubio}}, \bibinfo {author} {\bibfnamefont {A.~N.}\ \bibnamefont
  {Pasupathy}},\ and\ \bibinfo {author} {\bibfnamefont {C.~R.}\ \bibnamefont
  {Dean}},\ }\bibfield  {title} {\bibinfo {title} {Correlated electronic phases
  in twisted bilayer transition metal dichalcogenides},\ }\href
  {https://doi.org/10.1038/s41563-020-0708-6} {\bibfield  {journal} {\bibinfo
  {journal} {Nat. Mater.}\ }\textbf {\bibinfo {volume} {19}},\ \bibinfo {pages}
  {861} (\bibinfo {year} {2020})}\BibitemShut {NoStop}%
\bibitem [{\citenamefont {Wu}\ \emph {et~al.}(2018)\citenamefont {Wu},
  \citenamefont {Lovorn}, \citenamefont {Tutuc},\ and\ \citenamefont
  {MacDonald}}]{wuHubbardModelPhysics2018}%
  \BibitemOpen
  \bibfield  {author} {\bibinfo {author} {\bibfnamefont {F.}~\bibnamefont
  {Wu}}, \bibinfo {author} {\bibfnamefont {T.}~\bibnamefont {Lovorn}}, \bibinfo
  {author} {\bibfnamefont {E.}~\bibnamefont {Tutuc}},\ and\ \bibinfo {author}
  {\bibfnamefont {A.~H.}\ \bibnamefont {MacDonald}},\ }\bibfield  {title}
  {\bibinfo {title} {Hubbard {{Model Physics}} in {{Transition Metal
  Dichalcogenide Moir\'e Bands}}},\ }\href
  {https://doi.org/10.1103/PhysRevLett.121.026402} {\bibfield  {journal}
  {\bibinfo  {journal} {Phys. Rev. Lett.}\ }\textbf {\bibinfo {volume} {121}},\
  \bibinfo {pages} {026402} (\bibinfo {year} {2018})}\BibitemShut {NoStop}%
\bibitem [{\citenamefont {Wu}\ \emph {et~al.}(2019)\citenamefont {Wu},
  \citenamefont {Lovorn}, \citenamefont {Tutuc}, \citenamefont {Martin},\ and\
  \citenamefont {MacDonald}}]{wuTopologicalInsulatorsTwisted2019}%
  \BibitemOpen
  \bibfield  {author} {\bibinfo {author} {\bibfnamefont {F.}~\bibnamefont
  {Wu}}, \bibinfo {author} {\bibfnamefont {T.}~\bibnamefont {Lovorn}}, \bibinfo
  {author} {\bibfnamefont {E.}~\bibnamefont {Tutuc}}, \bibinfo {author}
  {\bibfnamefont {I.}~\bibnamefont {Martin}},\ and\ \bibinfo {author}
  {\bibfnamefont {A.~H.}\ \bibnamefont {MacDonald}},\ }\bibfield  {title}
  {\bibinfo {title} {Topological {{Insulators}} in {{Twisted Transition Metal
  Dichalcogenide Homobilayers}}},\ }\href
  {https://doi.org/10.1103/PhysRevLett.122.086402} {\bibfield  {journal}
  {\bibinfo  {journal} {Phys. Rev. Lett.}\ }\textbf {\bibinfo {volume} {122}},\
  \bibinfo {pages} {086402} (\bibinfo {year} {2019})}\BibitemShut {NoStop}%
\bibitem [{\citenamefont {Pan}\ \emph {et~al.}(2020{\natexlab{a}})\citenamefont
  {Pan}, \citenamefont {Wu},\ and\ \citenamefont
  {Das~Sarma}}]{panQuantumPhaseDiagram2020}%
  \BibitemOpen
  \bibfield  {author} {\bibinfo {author} {\bibfnamefont {H.}~\bibnamefont
  {Pan}}, \bibinfo {author} {\bibfnamefont {F.}~\bibnamefont {Wu}},\ and\
  \bibinfo {author} {\bibfnamefont {S.}~\bibnamefont {Das~Sarma}},\ }\bibfield
  {title} {\bibinfo {title} {Quantum phase diagram of a {{Moir\'e-Hubbard}}
  model},\ }\href {https://doi.org/10.1103/PhysRevB.102.201104} {\bibfield
  {journal} {\bibinfo  {journal} {Phys. Rev. B}\ }\textbf {\bibinfo {volume}
  {102}},\ \bibinfo {pages} {201104} (\bibinfo {year}
  {2020}{\natexlab{a}})}\BibitemShut {NoStop}%
\bibitem [{\citenamefont {Pan}\ \emph {et~al.}(2020{\natexlab{b}})\citenamefont
  {Pan}, \citenamefont {Wu},\ and\ \citenamefont
  {Das~Sarma}}]{panBandTopologyHubbard2020}%
  \BibitemOpen
  \bibfield  {author} {\bibinfo {author} {\bibfnamefont {H.}~\bibnamefont
  {Pan}}, \bibinfo {author} {\bibfnamefont {F.}~\bibnamefont {Wu}},\ and\
  \bibinfo {author} {\bibfnamefont {S.}~\bibnamefont {Das~Sarma}},\ }\bibfield
  {title} {\bibinfo {title} {Band topology, {{Hubbard}} model, {{Heisenberg}}
  model, and {{Dzyaloshinskii-Moriya}} interaction in twisted bilayer {{WSe}}
  2},\ }\href {https://doi.org/10.1103/PhysRevResearch.2.033087} {\bibfield
  {journal} {\bibinfo  {journal} {Phys. Rev. Research}\ }\textbf {\bibinfo
  {volume} {2}},\ \bibinfo {pages} {033087} (\bibinfo {year}
  {2020}{\natexlab{b}})}\BibitemShut {NoStop}%
\bibitem [{\citenamefont {Pan}\ and\ \citenamefont
  {Das~Sarma}(2021)}]{panInteractionDrivenFillingInducedMetalInsulator2021a}%
  \BibitemOpen
  \bibfield  {author} {\bibinfo {author} {\bibfnamefont {H.}~\bibnamefont
  {Pan}}\ and\ \bibinfo {author} {\bibfnamefont {S.}~\bibnamefont
  {Das~Sarma}},\ }\bibfield  {title} {\bibinfo {title} {Interaction-{{Driven
  Filling-Induced Metal-Insulator Transitions}} in {{2D Moir\'e Lattices}}},\
  }\href {https://doi.org/10.1103/PhysRevLett.127.096802} {\bibfield  {journal}
  {\bibinfo  {journal} {Phys. Rev. Lett.}\ }\textbf {\bibinfo {volume} {127}},\
  \bibinfo {pages} {096802} (\bibinfo {year} {2021})}\BibitemShut {NoStop}%
\bibitem [{\citenamefont {Pan}\ and\ \citenamefont
  {Das~Sarma}(2022)}]{panInteractionRangeTemperature2022}%
  \BibitemOpen
  \bibfield  {author} {\bibinfo {author} {\bibfnamefont {H.}~\bibnamefont
  {Pan}}\ and\ \bibinfo {author} {\bibfnamefont {S.}~\bibnamefont
  {Das~Sarma}},\ }\bibfield  {title} {\bibinfo {title} {Interaction range and
  temperature dependence of symmetry breaking in strongly correlated
  two-dimensional moir\'e transition metal dichalcogenide bilayers},\ }\href
  {https://doi.org/10.1103/PhysRevB.105.041109} {\bibfield  {journal} {\bibinfo
   {journal} {Phys. Rev. B}\ }\textbf {\bibinfo {volume} {105}},\ \bibinfo
  {pages} {041109} (\bibinfo {year} {2022})}\BibitemShut {NoStop}%
\bibitem [{\citenamefont {Ahn}\ and\ \citenamefont
  {Das~Sarma}(2022)}]{ahnDisorderinducedTwodimensionalMetalinsulator2022}%
  \BibitemOpen
  \bibfield  {author} {\bibinfo {author} {\bibfnamefont {S.}~\bibnamefont
  {Ahn}}\ and\ \bibinfo {author} {\bibfnamefont {S.}~\bibnamefont
  {Das~Sarma}},\ }\bibfield  {title} {\bibinfo {title} {Disorder-induced
  two-dimensional metal-insulator transition in moir\'e transition metal
  dichalcogenide multilayers},\ }\href
  {https://doi.org/10.1103/PhysRevB.105.115114} {\bibfield  {journal} {\bibinfo
   {journal} {Phys. Rev. B}\ }\textbf {\bibinfo {volume} {105}},\ \bibinfo
  {pages} {115114} (\bibinfo {year} {2022})}\BibitemShut {NoStop}%
\bibitem [{Mak({\natexlab{a}})}]{Mak}%
  \BibitemOpen
  \href@noop {} {\bibinfo {title} {K. {{Mak}}, private communications}}
  \BibitemShut {NoStop}%
\bibitem [{\citenamefont
  {Stern}(1967)}]{sternPolarizabilityTwoDimensionalElectron1967}%
  \BibitemOpen
  \bibfield  {author} {\bibinfo {author} {\bibfnamefont {F.}~\bibnamefont
  {Stern}},\ }\bibfield  {title} {\bibinfo {title} {Polarizability of a
  {{Two-Dimensional Electron Gas}}},\ }\href
  {https://doi.org/10.1103/PhysRevLett.18.546} {\bibfield  {journal} {\bibinfo
  {journal} {Phys. Rev. Lett.}\ }\textbf {\bibinfo {volume} {18}},\ \bibinfo
  {pages} {546} (\bibinfo {year} {1967})}\BibitemShut {NoStop}%
\bibitem [{\citenamefont {Das~Sarma}\ and\ \citenamefont
  {Hwang}(2015)}]{dassarmaScreeningTransport2D2015}%
  \BibitemOpen
  \bibfield  {author} {\bibinfo {author} {\bibfnamefont {S.}~\bibnamefont
  {Das~Sarma}}\ and\ \bibinfo {author} {\bibfnamefont {E.~H.}\ \bibnamefont
  {Hwang}},\ }\bibfield  {title} {\bibinfo {title} {Screening and transport in
  {{2D}} semiconductor systems at low temperatures},\ }\href
  {https://doi.org/10.1038/srep16655} {\bibfield  {journal} {\bibinfo
  {journal} {Sci Rep}\ }\textbf {\bibinfo {volume} {5}},\ \bibinfo {pages}
  {16655} (\bibinfo {year} {2015})}\BibitemShut {NoStop}%
\bibitem [{\citenamefont
  {Stern}(1980)}]{sternCalculatedTemperatureDependence1980}%
  \BibitemOpen
  \bibfield  {author} {\bibinfo {author} {\bibfnamefont {F.}~\bibnamefont
  {Stern}},\ }\bibfield  {title} {\bibinfo {title} {Calculated {{Temperature
  Dependence}} of {{Mobility}} in {{Silicon Inversion Layers}}},\ }\href
  {https://doi.org/10.1103/PhysRevLett.44.1469} {\bibfield  {journal} {\bibinfo
   {journal} {Phys. Rev. Lett.}\ }\textbf {\bibinfo {volume} {44}},\ \bibinfo
  {pages} {1469} (\bibinfo {year} {1980})}\BibitemShut {NoStop}%
\bibitem [{\citenamefont
  {Das~Sarma}(1986)}]{dassarmaTheoryFinitetemperatureScreening1986}%
  \BibitemOpen
  \bibfield  {author} {\bibinfo {author} {\bibfnamefont {S.}~\bibnamefont
  {Das~Sarma}},\ }\bibfield  {title} {\bibinfo {title} {Theory of
  finite-temperature screening in a disordered two-dimensional electron gas},\
  }\href {https://doi.org/10.1103/PhysRevB.33.5401} {\bibfield  {journal}
  {\bibinfo  {journal} {Phys. Rev. B}\ }\textbf {\bibinfo {volume} {33}},\
  \bibinfo {pages} {5401} (\bibinfo {year} {1986})}\BibitemShut {NoStop}%
\bibitem [{\citenamefont {Zala}\ \emph {et~al.}(2001)\citenamefont {Zala},
  \citenamefont {Narozhny},\ and\ \citenamefont
  {Aleiner}}]{zalaInteractionCorrectionsIntermediate2001}%
  \BibitemOpen
  \bibfield  {author} {\bibinfo {author} {\bibfnamefont {G.}~\bibnamefont
  {Zala}}, \bibinfo {author} {\bibfnamefont {B.~N.}\ \bibnamefont {Narozhny}},\
  and\ \bibinfo {author} {\bibfnamefont {I.~L.}\ \bibnamefont {Aleiner}},\
  }\bibfield  {title} {\bibinfo {title} {Interaction corrections at
  intermediate temperatures: {{Longitudinal}} conductivity and kinetic
  equation},\ }\href {https://doi.org/10.1103/PhysRevB.64.214204} {\bibfield
  {journal} {\bibinfo  {journal} {Phys. Rev. B}\ }\textbf {\bibinfo {volume}
  {64}},\ \bibinfo {pages} {214204} (\bibinfo {year} {2001})}\BibitemShut
  {NoStop}%
\bibitem [{\citenamefont {Buterakos}\ \emph {et~al.}(2021)\citenamefont
  {Buterakos}, \citenamefont {Vu}, \citenamefont {Yu},\ and\ \citenamefont
  {Das~Sarma}}]{buterakosPresenceAbsenceTwodimensional2021}%
  \BibitemOpen
  \bibfield  {author} {\bibinfo {author} {\bibfnamefont {D.}~\bibnamefont
  {Buterakos}}, \bibinfo {author} {\bibfnamefont {D.}~\bibnamefont {Vu}},
  \bibinfo {author} {\bibfnamefont {J.}~\bibnamefont {Yu}},\ and\ \bibinfo
  {author} {\bibfnamefont {S.}~\bibnamefont {Das~Sarma}},\ }\bibfield  {title}
  {\bibinfo {title} {Presence versus absence of two-dimensional {{Fermi}}
  surface anomalies},\ }\href {https://doi.org/10.1103/PhysRevB.103.205154}
  {\bibfield  {journal} {\bibinfo  {journal} {Phys. Rev. B}\ }\textbf {\bibinfo
  {volume} {103}},\ \bibinfo {pages} {205154} (\bibinfo {year}
  {2021})}\BibitemShut {NoStop}%
\bibitem [{Mak({\natexlab{b}})}]{Mak2}%
  \BibitemOpen
  \href@noop {} {\bibinfo {title} {{{T}}. {{L}}i, {{J}}. {{S}}han, and {{K}}.
  {{Mak}}, private communication and unpublished}}\BibitemShut
  {NoStop}%
\bibitem [{\citenamefont {Hartnoll}\ and\ \citenamefont
  {Mackenzie}(2021)}]{hartnollPlanckianDissipationMetals2021}%
  \BibitemOpen
  \bibfield  {author} {\bibinfo {author} {\bibfnamefont {S.~A.}\ \bibnamefont
  {Hartnoll}}\ and\ \bibinfo {author} {\bibfnamefont {A.~P.}\ \bibnamefont
  {Mackenzie}},\ }\bibfield  {title} {\bibinfo {title} {Planckian
  {{Dissipation}} in {{Metals}}},\ }\href {http://arxiv.org/abs/2107.07802}
  {\bibfield  {journal} {\bibinfo  {journal} {arXiv:2107.07802}\ } (\bibinfo
  {year} {2021})}\BibitemShut {NoStop}%
\bibitem [{\citenamefont {Ahn}\ and\ \citenamefont
  {Sarma}(2022)}]{ahnPlanckianProperties2D2022}%
  \BibitemOpen
  \bibfield  {author} {\bibinfo {author} {\bibfnamefont {S.}~\bibnamefont
  {Ahn}}\ and\ \bibinfo {author} {\bibfnamefont {S.~D.}\ \bibnamefont
  {Sarma}},\ }\bibfield  {title} {\bibinfo {title} {Planckian properties of
  {{2D}} semiconductor systems},\ }\href {http://arxiv.org/abs/2204.02982}
  {\bibfield  {journal} {\bibinfo  {journal} {arXiv:2204.02982}\ } (\bibinfo
  {year} {2022})}\BibitemShut {NoStop}%
\bibitem [{\citenamefont {Das~Sarma}\ and\ \citenamefont
  {Hwang}(2013)}]{dassarmaUniversalDensityScaling2013}%
  \BibitemOpen
  \bibfield  {author} {\bibinfo {author} {\bibfnamefont {S.}~\bibnamefont
  {Das~Sarma}}\ and\ \bibinfo {author} {\bibfnamefont {E.~H.}\ \bibnamefont
  {Hwang}},\ }\bibfield  {title} {\bibinfo {title} {Universal density scaling
  of disorder-limited low-temperature conductivity in high-mobility
  two-dimensional systems},\ }\href
  {https://doi.org/10.1103/PhysRevB.88.035439} {\bibfield  {journal} {\bibinfo
  {journal} {Phys. Rev. B}\ }\textbf {\bibinfo {volume} {88}},\ \bibinfo
  {pages} {035439} (\bibinfo {year} {2013})}\BibitemShut {NoStop}%
\bibitem [{com()}]{comm}%
  \BibitemOpen
  \href@noop {} {\bibinfo {title} {We have been informed by {{A}}.
  {{P}}asupathy that indeed recent experiments in {{C}}olumbia {{U}}niversity
  using purer samples observe a larger metallic regime, and we predict that the
  reverse would happen in dirtier samples}}\BibitemShut {NoStop}%
\end{thebibliography}

%

\end{document}